\documentclass[12pt]{article}
\usepackage[centertags]{amsmath}
\usepackage{amsfonts}
\usepackage{amssymb}
\usepackage{amsthm}
\usepackage{newlfont}
\usepackage{stmaryrd}
\usepackage{mathrsfs}
\usepackage{pstricks,pst-node}
\usepackage{palatino}  
\usepackage{fancyhdr}
\usepackage{graphicx}
\usepackage{fancybox}
\usepackage{multibox}
\usepackage{geometry}
\usepackage[dvips]{epsfig}
\usepackage{eepic}
\usepackage{pst-plot}

  \let\La=\Lambda

\newcommand{\opunit}{\text{1}\kern-0.22em\text{l}}

\newcommand{\bde}{\begin{definition}}
\newcommand{\ede}{\end{definition}}
\newcommand{\beq}{\begin{equation}}
\newcommand{\eeq}{\end{equation}}
\newcommand{\ben}{\begin{enumerate}}
\newcommand{\een}{\end{enumerate}}
\newcommand{\ble}{\begin{lemma}}
\newcommand{\ele}{\end{lemma}}
\newcommand{\bpr}{\begin{proof}}
\newcommand{\epr}{\end{proof}}

\title{First-order transitions for very nonlinear sigma models.}

\author{
  {\normalsize Aernout C.~D.~van Enter}        \\[-1mm]
  {\normalsize\it Centre for Theoretical Physics}   \\[-1.5mm]
  {\normalsize\it Rijksuniversiteit Groningen}         \\[-1.5mm]
  {\normalsize\it Nijenborgh 4}                \\[-1.5mm]
  {\normalsize\it 9747 AG Groningen}           \\[-1.5mm]
  {\normalsize\it THE NETHERLANDS}             \\[-1mm]
  {\normalsize\tt aenter@phys.rug.nl}        \\[-1mm]
\\ [-1mm]
  {\normalsize Senya B.~Shlosman}            \\[-1mm]
  {\normalsize\it Centre de  Physique Theorique}   \\[-1.5mm]
  {\normalsize\it CNRS}         \\[-1.5mm]
  {\normalsize\it Luminy, Case 907}                \\[-1.5mm]
  {\normalsize\it F13288, Marseille, Cedex 9}           \\[-1.5mm]
  {\normalsize\it FRANCE}             \\[-1mm]
  {\normalsize\tt shlosman@cpt.univ-mrs.fr}        \\[-1mm]
{\protect\makebox[5in]{\quad}}}
\begin{document}
\maketitle \baselineskip=14pt \noindent {\bf Abstract.} In this
contribution we discuss the occurrence of first-order transitions in 
temperature in various short-range lattice models with a rotation symmetry. 
Such transitions turn out to be widespread under the condition 
that the interaction potentials are sufficiently nonlinear.

\bigskip

\noindent{\em In memory of John Lewis, who by his kind guidance and 
excellent scientific and professional sense has always been a 
stimulus and an inspiration.}

\newtheorem{theorem}{Theorem}          
\newtheorem{lemma}{Lemma}              
\newtheorem{proposition}[lemma]{Proposition}
\newtheorem{corollary}[lemma]{Corollary}
\newtheorem{definition}[theorem]{Definition}
\newtheorem{conjecture}[theorem]{Conjecture}
\newtheorem{claim}[theorem]{Claim}
\newtheorem{observation}[theorem]{Observation}
\def\proof{\par\noindent{\it Proof.\ }}
\def\reff#1{(\ref{#1})}

\let\zed=\bbbz 
\let\szed=\bbbz 
\let\IR=\bbbr 
\let\R=\bbbr 
\let\sIR=\bbbr 
\let\IN=\bbbn 
\let\IC=\bbbc 

\def\nl{\medskip\par\noindent}

\def\scrb{{\cal B}}
\def\scrg{{\cal G}}
\def\scrf{{\cal F}}
\def\scrl{{\cal L}}
\def\scrr{{\cal R}}
\def\scrt{{\cal T}}
\def\pfin{{\cal S}}
\def\prob{M_{+1}}
\def\cql{C_{\rm ql}}
\def\bydef{\stackrel{\rm def}{=}}   
\def\qed{\hbox{\hskip 1cm\vrule width6pt height7pt depth1pt \hskip1pt}\bigskip}
\def\remark{\medskip\par\noindent{\bf Remark:}}
\def\remarks{\medskip\par\noindent{\bf Remarks:}}
\def\example{\medskip\par\noindent{\bf Example:}}
\def\examples{\medskip\par\noindent{\bf Examples:}}
\def\nonexamples{\medskip\par\noindent{\bf Non-examples:}}

\newenvironment{scarray}{
          \textfont0=\scriptfont0
          \scriptfont0=\scriptscriptfont0
          \textfont1=\scriptfont1
          \scriptfont1=\scriptscriptfont1
          \textfont2=\scriptfont2
          \scriptfont2=\scriptscriptfont2
          \textfont3=\scriptfont3
          \scriptfont3=\scriptscriptfont3
        \renewcommand{\arraystretch}{0.7}
        \begin{array}{c}}{\end{array}}

\def\wspec{w'_{\rm special}}
\def\mup{\widehat\mu^+}
\def\mupm{\widehat\mu^{+|-_\Lambda}}
\def\pip{\widehat\pi^+}
\def\pipm{\widehat\pi^{+|-_\La\bibitem{mi}

mbda}}
\def\ind{{\rm I}}
\def\const{{\rm const}}

\bibliographystyle{plain}


\maketitle

\section{Introduction}

One of the main predictions of the Renormalisation Group (RG) theory is what
is called \textquotedblleft universality\textquotedblright. This means that in
great generality the nature of the phase transition between high-temperature
and low-temperature phases and the corresponding critical exponents depend
only on dimension, symmetry and the range of the interaction. Here
\textit{range} means short-range (finite-range or sufficiently fast decaying
with distance) or long-range (slowly decaying at large distances). The
classical Landau mean-field theory similarly predicts that the nature of the
spontaneously broken symmetry determines the order of the transition. Although
in many cases such RG predictions have been confirmed, there are some examples
where, somewhat unexpectedly, first-order instead of the predicted
second-order (or absence of any) transitions were observed, see e.g.
\cite{BisChaCra,KunZum1,SokSta, SonTch}.

In some cases, such as the nearest-neighbour $q$-state Potts models, one might
think that it is the nature of the broken (permutation) symmetry which governs
whether there is a first-order (at high $q$) or a second-order (at low $q$)
transition. But the generalization of this statement is hard to make.  Indeed,
as Onsager \cite{Fey} already knew, there seems to be no general method to
predict whether a transition is first-order or second-order. For example, it
was shown in \cite{gobmer} (which extended the related work of
\cite{BisChaCra}) that a 3-state Potts model in dimension two, with an
interaction of finite but large range, has a first-order transition, while for
the nearest-neighbour model the transition is of second order.

In this contribution, we review results from \cite{ES1, ES2} where we
exhibited a different class of models.  Though they possess global rotation
symmetries, they undergo first-order transitions, whereas the universality
predictions of the RG  suggest second-order transitions. More precisely, for
short-range ferromagnetic, $d$-dimensional, rotation-invariant $n$-vector
models standard lore \cite{convention} (which as we here show can be violated)
predicts the following (\textquotedblleft universal\textquotedblright) behaviour:

\begin{itemize}
\item \smallskip\ If $d=2$, there is a unique Gibbs measure at any positive
temperature. For $n=2$ \noindent(classical XY spins -- or the
\textquotedblleft nonlinear sigma model\textquotedblright\ in field
theoretical language)  there is nevertheless an infinite-order transition
between a low-temperature Kosterlitz-Thouless phase with slowly decaying
correlations and a high-temperature phase with exponential correlation decay.
For higher $n$ there is no phase transition.

\item If $d=3$ or higher, there is a second-order transition between a
magnetized low-temperature phase and a high-temperature phase. If $d=3$, one
has $n$-dependent critical exponents, in higher dimensions one obtains
mean-field exponents.
\end{itemize}

\smallskip\  

Below we present a rather wide class of models in which this standard lore is violated.

\smallskip

\section{ Notation and some background}

For general background on the theory of Gibbs measures we refer to \cite{EFS,
Geo, GeoHagMae, Sin, Sim}. We will consider spin models defined on the lattice
$\mathbb{Z}^{d}$ with spins taking values on the $n$-dimensional unit sphere.

We will use small Greek letters $\sigma,\eta,\ldots$ to denote spin
configurations in finite or infinite boxes. The nearest-neighbour Hamiltonians
in a box $\Lambda$, for which we take  a $d$-dimensional torus, will be given by%

\begin{equation}
H^{\Lambda}(\sigma)=\sum_{\langle i,j\rangle\subset\Lambda}U(\sigma_{i}%
\cdot\sigma_{j})+\sum_{i\in\Lambda}h\cdot\sigma_{i}.
\end{equation}
Associated to these Hamiltonians $H^{\Lambda}(\sigma)$ are Gibbs measures
\begin{equation}
\mu^{\Lambda}(d\sigma)={\frac{1}{Z^{\Lambda}}}\exp[-H^{\Lambda}(\sigma
)]\mu_{0}^{\Lambda}(d\sigma)
\end{equation}

Here $\mu_{0}^{\Lambda}(d\sigma)$ denotes the rotation-invariant product
measure. The choice of the function $U$ of the inner product between the spins
at neighbouring sites will determine our model. We will study only the case of
nearest-neighbour interactions. The reason is that our method is based on the
use of certain correlation inequalities, called chessboard estimates, see
below. These inequalities hold once the measures $\mu^{\Lambda}$ have the
\emph{Reflection Positivity (RP) }property, see again below, which RP holds
for the n.n. interactions, see \cite{Geo,RP}.

\noindent The choice
\begin{equation}
U(x)=-x
\end{equation}

\noindent provides the standard classical XY and Heisenberg models.
Equivalently, as a function $W$ of the difference angle $\theta$ between
neighbouring spins, this means choosing
\begin{equation}
W(\theta)= U(cos (\theta))= -cos(\theta)
\end{equation}

\section{Results, and some remarks on proofs.}

In this section we describe our results on the nonstandard $n$-vector models.

We start with the case of zero field $h=0$ and $d=2,n=2$, so we have classical
$XY$ spins in two dimensions. This seems to be the first case which was
considered in the literature \cite{DomSchSve} as an example of the phenomenon
we display. However, the arguments in that paper -- which we here prove to be
correct -- were later contested \cite{Kno,vHim}. The original choice of
\cite{DomSchSve} was%

\begin{equation}
W(\theta) = -{({\frac{1 }{2}}(1 + cos \theta))}^{p}%
\end{equation}

\noindent with $p$ large enough. A simpler but essentially similar model was
introduced in \cite{AC}:%

\begin{equation}
W(\theta)=\left\{
\begin{array}
[c]{cc}%
-1 & \text{ if }\left\vert \theta\right\vert <\varepsilon,\\
0 & \text{ otherwise,}%
\end{array}
\right.
\end{equation}

\noindent where the parameter $\varepsilon$ is small enough.

Both these potentials have the form of a deep (depth $=1$) and narrow (width
$\sim\frac{1}{\sqrt{p}},$  cf width $=2\varepsilon$) well, compared to the
standard, rather shallow-well, cosine shape.

The second model is a square well (or a top hat) potential, in which the
distinction between being in or out of the well is unambiguous, in the first
model there is a slight arbitrariness, and one has to make a choice to fix it.

First we notice that by the Mermin-Wagner theorem \cite{MerWag,DS2,IofShlVel,
Pfi}, all Gibbs measures are rotation-invariant, so that the spontaneous
magnetisation is necessarily zero. This does not prevent, however, the
presence of multiple Gibbs measures (as was known already from the model of
\cite{Shl2}, where a discrete, chiral, symmetry was shown to be broken).

Our first result is about the square-well model:

\smallskip

\noindent\textbf{Theorem}

\noindent\emph{For $\varepsilon$ small enough, there is a transition
temperature where two Gibbs measures, an ``ordered'' one and a ``disordered''
one, coexist. In the ordered state most bonds are ordered, in the sense that
the two spins at its ends have a difference angle smaller than $2 \varepsilon$
(they are in the well), in the disordered state the opposite is true.}

\smallskip

\textbf{Remarks about the proof}: The proof is a fairly straightforward
application of the Reflection Positivity, chessboard estimates method, which
was developed by Dyson, Fr\"{o}hlich, Israel, Lieb and Simon \cite{RP}. For
the benefit of the reader we recall these concepts.

Let $R_{L}$ be a reflection of our torus $\Lambda$ in some plane $L$ passing
through its sites. (To have such a symmetry plane, $\Lambda$ has to be of even
size.) Then $L$ cuts $\Lambda$ into two halves, $\Lambda_{+}$ and $\Lambda
_{-},$ so that $R_{L}\left(  \Lambda_{\pm}\right)  =\Lambda_{\mp}.$ Let
$\Lambda_{0}=\Lambda\cap L;$ in the case $d=2$ this intersection is a pair of
meridians of $\Lambda.$ Let us consider the conditional distribution of the
measure $\mu^{\Lambda}$ under condition that the restriction $\sigma
_{\Lambda_{0}}\equiv\sigma\Bigm|_{\Lambda_{0}}$ is fixed. Then it is easy to
see that for every value of $\sigma_{\Lambda_{0}}$ the conditional measure
$\mu^{\Lambda}\left(  \cdot\Bigm|\sigma_{\Lambda_{0}}\right)  $ splits into a
product of two identical measures, $\mu^{\Lambda_{\pm},\sigma_{\Lambda_{0}}}$,
living on corresponding halves, with $R_{L}\left(  \mu^{\Lambda_{\pm}%
,\sigma_{\Lambda_{0}}}\right)  =\mu^{\Lambda_{\mp},\sigma_{\Lambda_{0}}}.$
From that it follows immediately that for \textit{every} function $C\left(
\sigma_{\Lambda}\right)  =C\left(  \sigma_{\Lambda_{+}}\right)  ,$
\textit{depending only on the \textquotedblleft left\textquotedblright%
\ variables} $\sigma_{\Lambda_{+}}$, we have %

\begin{equation}
\int C~R_{L}C~d\mu^{\Lambda}\geq0.
\end{equation}
The Reflection Positivity property is precisely the validity of this
inequality. The details can be found in \cite{Geo}, Theorem 17.21. Note that
one can choose the symmetry plane $L$ arbitrary, so in fact we have many such
inequalities, and one corollary of this set of inequalities is the following
\textit{chessboard estimate}. 

To describe the simplest example of such an estimate let us consider a random
variable $D\left(  \sigma_{\Lambda}\right)  ,$ which depends on just one spin
value, $\sigma_{0},$ where $0\in\Lambda$ is the origin. Then for its expected
value $\left\langle D\left(  \sigma_{0}\right)  \right\rangle _{\mu^{\Lambda}%
}\equiv\int D\left(  \sigma_{0}\right)  ~d\mu^{\Lambda}\left(  \sigma
_{\Lambda}\right)  $ we have
\begin{equation}
\left\langle D\left(  \sigma_{0}\right)  \right\rangle _{\mu^{\Lambda}}%
\leq\left[  \left\langle \prod_{\substack{x\in\Lambda,\\x\text{ is even}%
}}D\left(  \sigma_{x}\right)  \right\rangle _{\mu^{\Lambda}}\right]
^{\frac{4}{\left\vert \Lambda\right\vert }}.\label{chess}%
\end{equation}
Here $D\left(  \sigma_{x}\right)  $ is the same function $D$, computed at
value $\sigma_{x},$ and we take a product over all sites $x$ with both
coordinates even. Note that if the interaction $U$ is identically zero, then
the measure  $\mu^{\Lambda}$ is just the product measure $\mu_{0}^{\Lambda
}(d\sigma),$ and the last inequality becomes an equality.

Applying this kind of chessboard estimate to various observables it is fairly
straightforward to show that at low temperatures most bonds are ordered and
that at high temperatures most bonds are disordered. The main step then left
is to prove that uniformly for all temperatures in a temperature interval,
including both high and low temperatures, the probability for two arbitrary
bonds to be different (one ordered, one disordered) is small. Indeed, the only
way these two properties can hold simultaneously is the existence of an
intermediate temperature at which both the ordered and the disordered phase coexist.

We will explain now how an estimate of the probability that a certain bond
$b_{1}$ is ordered, while another one, $b_{2},$ is disordered, can be
obtained. If such an event happens, then there exists a contour $\gamma,$
separating $b_{1}$ and $b_{2},$ formed by sites which have a pair of
orthogonal bonds, one of them being ordered and another disordered. For
example, one obtains such a contour by taking the appropriate component of the
boundary of the set of ordered bonds, containing the bond $b_{1}.$ Therefore
it is enough to obtain a Peierls-type contour estimate, which shows that long
contours are improbable. More precisely, we need to show that the probability
for the occurrence of a contour $\gamma$ of size $|\gamma|$ is exponentially
small in $|\gamma|$. 

By using again the chessboard estimate, it is possible to show that it is
enough to obtain the desired estimate for just one single contour, $\Gamma,$
called the \textit{universal }contour. This universal contour contains
\textit{all }bonds of $\Lambda;$ half of them are ordered, and the remaining
half - disordered. In our case the event $\Gamma$ happens iff all the bonds
adjacent to sites $x$ with $x_{1}+x_{2}=0\operatorname{mod}4$ are ordered,
while those adjacent to sites with $x_{1}+x_{2}=2\operatorname{mod}4$ are
disordered. Since the size $\left\vert \Gamma\right\vert =\left\vert
\Lambda\right\vert ,$ we have to show that
\begin{equation}
\mathbf{\Pr}\left(  \Gamma\right)  \leq\exp\left\{  -c\left\vert
\Lambda\right\vert \right\}  \label{uc}%
\end{equation}
with the constant $c$ sufficiently large. (The concept of a "universal
contour" goes back to the pioneering paper by Fr\"{o}hlich and Lieb \cite{FL}.

The remaining computations go just as in the proof for the large-$q$ Potts
model, given in detail in \cite{KS} or \cite{Shl1}. This similarity with the
Potts problem was already remarked upon in \cite{DomSchSve}. The
correspondence is that $\varepsilon$, (and the same holds for $\frac{1}%
{\sqrt{p}}$), plays the role of the small parameter $\frac{1}{q}$. To estimate
the probability of the universal contour, one has to integrate over all
configurations such that the prescribed arrangement of ordered and disordered
bonds occurs. To do it we observe that the total partition function of an
$N$-by-$N$ square satisfies
\begin{equation}
Z_{N}\geq\max[1,{({\frac{1}{2\pi}}\varepsilon\ exp2\beta)}^{N^{2}}].
\end{equation}
Indeed, on the one hand we use that the potential is positive, and on the
other hand we get a lower bound by taking the integral at each site over the
interval $[-\frac{1}{2}\varepsilon,\frac{1}{2}\varepsilon]$, so that each bond
is ordered.

The restricted partition function $Z_{\Gamma},$ which is the integral over all
configurations compatible with the universal contour satisfies
\begin{equation}
Z_{\Gamma}\leq{(2\varepsilon)}^{\frac{1}{4}N^{2}}\exp\left\{  \frac{\beta
N^{2}}{2}+O(N)\right\}  .
\end{equation}
From these two estimates the bound $\left(  \ref{uc}\right)  $ follows.

The final conclusion is now, that somewhere inside our temperature interval
there is a value $\beta_{t},$ at which a first-order transition happens
between an ordered and a disordered Gibbs measure, as was first numerically
found in \cite{DomSchSve}. The value of $\beta_{t}$ is approximately given by
$2\beta_{t}=-\ln\varepsilon$.

The non-square-well model can be treated in a very similar way (see
\cite{ES1,ES2} and also \cite{BiKo}). The ordered Gibbs measure has a
polynomial spin-spin correlation decay of Kosterlitz-Thouless type.

\bigskip

\textbf{Generalisations:}

1) The same method of proof works if either the spin dimensionality $n$ or the
dimension $d$ of the lattice is larger than $2$ (or both). For the case of
Heisenberg spins ($n=3$) in $d=2$, the first-order transition was first found
numerically 
\cite{BloGuoHil, BloGuo}. In this case
presumably both the low-temperature and the high-temperature phase have
exponential decay of the spin-spin correlations. For the $n\rightarrow\infty$
spherical limit see also \cite{CarPel, CarMoPel}.

For $d \geq3$, the Mermin-Wagner theorem does not apply anymore, and the
low-temperature phase now displays a spontaneous magnetisation.

\smallskip

2) In a small external field there still is a first-order transition between
an ordered (strongly magnetised in the direction of the field) 
and a disordered (weakly magnetised in the direction of the field) 
phase, which now we expect to be both pure phases
(extremal Gibbs measures) also in higher dimensions. 

\smallskip

3) Instead of a single well, one can also consider potentials having the shape
of repeated wells in wells (or a hat-in-a-hat-in-a-hat...), which give rise to
possibly infinitely many transitions. A choice of such a Seuss (\cite{Seu})
potential is $U(x) = - \sum_{n} 2^{-n} 1_{\varepsilon_{n}}(x)$ with
${\varepsilon_{n}}(= \varepsilon_{n-1}^{3})= \varepsilon^{3^{n-1}}$, with the
first $\varepsilon$ small enough. Such transitions in which one keeps jumping
in deeper and deeper wells, can occur either between
nonmagnetised-nonmagnetised, magnetised-nonmagnetised or magnetised-magnetised
Gibbs measures, and the nonmagnetised measures may display either exponential
or Kosterlitz-Thouless decay.

\smallskip

4) Instead of ferromagnetic models the argument also works for nematic liquid
crystal $RP^{n}$ models \cite{Las, LasLeb}, in which one considers
interactions of the form $U(x)=-x^{2p}$ for which there are two minima on the
interval $[-1,+1]$. Here even for $p=1$, and $n=3$, first-order transitions
were found numerically (see e.g. \cite{KunZum1,KunZum2,Ro}) in $d=3$, whereas
the occurrence of a transition in the limit $n \to\infty$ in $d=2$ has been a
matter of controversy \cite{SokSta, SonTch}. Additional numerical references
are mentioned in \cite{ES2}. Further models of this type, with a larger number
of sharp minima, combined with a term causing a chiral symmetry breaking as in
\cite{Shl2}, are considered in \cite{MeNa}.

\smallskip

5) Similarly as for the Potts gauge model of \cite{KS}, we can prove the
existence of a first-order transition in various nonlinear lattice gauge
models with continuous symmetries. In some of these models first-order
transitions were initially concluded on the basis of numerical data (see e.g.
\cite{EspTag,ABLS,Smi}). In \cite{ES2} we provide the first occasion where a
first-order transition for a lattice gauge model in the presence of a
continuous symmetry can be proven.

\smallskip

6) Instead of the above spin systems with a compact rotation symmetry, one can
also consider continuous unbounded-spin systems which possess a non-compact
symmetry (that is, they are ``massless''). In this case the symmetry describes
a shift in the height (average) of the spin. Again a similar construction
works \cite{BisKot1}, now showing coexistence of ``gradient Gibbs measures''.

\smallskip

7) For quantum spin systems the ingenious analysis of \cite{BisChaSta} shows
that again a first-order transition occurs, once the potential is sufficiently
nonlinear and the spin number is large enough.

\section{Conclusions and Comments}

In many cases, some of them also of direct physical interest, first-order
transitions occur instead of the second-order transition, which a naive
universality argument would predict. Such first-order transitions tend to
occur more easily when either the nonlinearity parameter $p$, the
spin-dimensionality $n$  (the spherical limit), or the dimension $d$ of the
lattice  (the mean-field limit,   \cite{BisCha})  is large.

In this paper we have presented a number of models where we can prove these
transitions for sufficiently strong nonlinearities. Furthermore, if one adds
an external field, the transition persists. Although the two phases on both
sides of the transition can have different characters (breaking other
symmetries, having polynomial or exponential decay of the spin-spin
correlations, etc), it can be shown that no ``intermediate'' phases exist
(there is a ``forbidden gap'' for the energy variable \cite{BiKo}).

The method of Reflection Positivity we use here has the disadvantage that one
is limited in the interactions one can take, in the sense that they need to be
defined on the unit cube. As compared to the more robust Pirogov-Sinai contour
methods, RP methods have the advantage, however, that they are more generally
applicable in that we need no information about the phases on both sides of
the transition. For Pirogov-Sinai methods to work, however, the phases need to
be pure and will typically have some kind of exponential decay of
correlations. On the other hand, with RP methods it has not been possible up
to now to obtain results for surface or interface properties or about the
completeness of the phase diagram.

If one varies the nonlinearity parameter in $d=2$, one moves towards a
critical point where the second-order transition is expected to have Ising
characteristics, in higher dimensions varying the nonlinearity parameter will
lead one to a tricitical point.

Although our proofs require a fairly large nonlinearity (that is, a large
value of $p$, or a small value of $\varepsilon$), either by numerical methods
or in the large-$n$ or large-$d$ limit we expect, and sometimes know, that the
type of first-order transitions we studied will occur for much smaller values
of $p$, $n$, and{/}or $d$, especially for the liquid-crystal models.

\bigskip

\noindent\emph{Acknowledgements} During this work we had very stimulating
input from various colleagues, in particular L.~Chayes, who developed
independently a number of these ideas, and E.~Domany and A.~Schwimmer who
urged us to also consider lattice gauge models. Some useful remarks on the
manuscript by C.~K\"{u}lske are gratefully acknowledged.







\end{document}